\documentclass[12pt,hyper]{JHEP3}
\usepackage{graphics,psfrag}
\usepackage{amsmath,amsthm,amssymb,epsfig,euscript,array,cite,cancel}
\usepackage{cases,empheq}
\theoremstyle{definition}

\theoremstyle{remark}

\newcounter{multieqs}

\newcommand{\be}{\begin{equation}}
\newcommand{\ee}{\end{equation}}
\newcommand{\eq}[1]{(\ref{#1})}
\newcommand{\bit}{\begin{itemize}}  \newcommand{\eit}{\end{itemize}}

\newcommand{\bm}[1]{\mbox{\boldmath $#1$}}
\newcommand{\rf}[1]{(\ref{#1})}

\def\bd{\begin{document}}
\def\ed{\end{document}}
\def\nn{\nonumber}
\def\bea{\begin{eqnarray}}
\def\eea{\end{eqnarray}}
\let\bm=\bibitem

\def\la{\langle}
\def\ra{\rangle}

\def\npb#1#2#3{Nucl. Phys. {\bf{B#1}} #3 (#2)}
\def\plb#1#2#3{Phys. Lett. {\bf{#1B}} #3 (#2)}
\def\prl#1#2#3{Phys. Rev. Lett. {\bf{#1}} #3 (#2)}
\def\prd#1#2#3{Phys. Rev. {D \bf{#1}} #3 (#2)}
\def\cmp#1#2#3{Comm. Math. Phys. {\bf{#1}} #3 (#2)}
\def\cqg#1#2#3{Class. Quantum Grav. {\bf{#1}} #3 (#2)}
\def\nppsa#1#2#3{Nucl. Phys. B (Proc. Suppl.) {\bf{#1A}}#3 (#2)}
\def\ap#1#2#3{Ann. of Phys. {\bf{#1}} #3 (#2)}
\def\ijmp#1#2#3{Int. J. Mod. Phys. {\bf{A#1}} #3 (#2)}
\def\rmp#1#2#3{Rev. Mod. Phys. {\bf{#1}} #3 (#2)}
\def\mpla#1#2#3{Mod. Phys. Lett. {\bf A#1} #3 (#2)}
\def\jhep#1#2#3{J. High Energy Phys. {\bf #1} #3 (#2)}
\def\atmp#1#2#3{Adv. Theor. Math. Phys. {\bf #1} #3 (#2)}

\def\N{{\cal N}}
\def\sst{\scriptscriptstyle}
\def\thetabar{\bar\theta}
\def\Tr{{\rm Tr}}
\def\one{\mbox{1 \kern-.59em {\rm l}}}

\def\a{\alpha}      \def\da{{\dot\alpha}}  \def\dA{{\dot A}}
\def\b{\beta}       \def\db{{\dot\beta}}  
\def\g{\gamma}  \def\G{\Gamma}  \def\dc{{\dot\gamma}}  
\def\d{\delta}  \def\D{\Delta}  \def\ddt{\dot\delta}  
\def\e{\epsilon}        \def\ve{\varepsilon}  
\def\f{\phi}    \def\F{\Phi}    \def\vvf{\f}  
\def\h{\eta}  
\def\k{\kappa}  
\def\l{\lambda} \def\L{\Lambda}  
\def\m{\mu} \def\n{\nu}  
\def\o{\omega}  
\def\p{\pi} \def\P{\Pi}  
\def\r{\rho}  
\def\s{\sigma}  \def\S{\Sigma}  
\def\t{\tau}  
\def\th{\theta} \def\Th{\Theta} \def\vth{\vartheta}  
\def\X{\Xeta}  
\def\z{\zeta}  

\def\na{\nabla}  

\def\cA{{\cal A}} \def\cB{{\cal B}} \def\cC{{\cal C}}  
\def\cD{{\cal D}} \def\cE{{\cal E}} \def\cF{{\cal F}}  
\def\cG{{\cal G}} \def\cH{{\cal H}} \def\cI{{\cal I}}  
\def\cJ{{\cal J}} \def\cK{{\cal K}} \def\cL{{\cal L}}  
\def\cM{{\cal M}} \def\cN{{\cal N}} \def\cO{{\cal O}}  
\def\cP{{\cal P}} \def\cQ{{\cal Q}} \def\cR{{\cal R}}  
\def\cS{{\cal S}} \def\cT{{\cal T}} \def\cU{{\cal U}}  
\def\cV{{\cal V}} \def\cW{{\cal W}} \def\cX{{\cal X}}  
\def\cY{{\cal Y}} \def\cZ{{\cal Z}}

\def\ua{\underline{\alpha}}  
\def\uc{\underline{\phantom{\alpha}}\!\!\!\gamma}  
\def\um{\underline{\mu}}  
\def\ud{\underline\delta}  
\def\ue{\underline\epsilon}  
\def\una{\underline a}\def\unA{\underline A}  
\def\unb{\underline b}\def\unB{\underline B}  
\def\unc{\underline c}\def\unC{\underline C}  
\def\und{\underline d}\def\unD{\underline D}  
\def\une{\underline e}\def\unE{\underline E}  
\def\unf{\underline{\phantom{e}}\!\!\!\! f}\def\unF{\underline F}  
\def\unm{\underline m}\def\unM{\underline M}  
\def\unn{\underline n}\def\unN{\underline N}  
\def\unp{\underline{\phantom{a}}\!\!\! p}\def\unP{\underline P}  
\def\unq{\underline{\phantom{a}}\!\!\! q}  
\def\unQ{\underline{\phantom{A}}\!\!\!\! Q}  
\def\unH{\underline{H}}  
  
\def\As {{A \hspace{-6.4pt} \slash}\;}  
\def\bs {{b \hspace{-6.4pt} \slash}\;}  
\def\Ds {{D \hspace{-6.4pt} \slash}\;}
\def\Gts {{\Gt \hspace{-6.4pt} \slash}\;}
\def\ds {{\del \hspace{-6.4pt} \slash}\;}  
\def\ss {{\s \hspace{-6.4pt} \slash}\;}  
\def\ks {{ k \hspace{-6.4pt} \slash}\;}  
\def\ps {{p \hspace{-6.4pt} \slash}\;}   
\def\xs {{x \hspace{-6.4pt} \slash}\;}  
\def\pas {{{p_1} \hspace{-6.4pt} \slash}\;}  
\def\pbs {{{p_2} \hspace{-6.4pt} \slash}\;}   
\def\cFs {{{\cal F} \hspace{-6.4pt} \slash}\;}

\def\Ah{{\hat{A}}}  
\def\Dh{{\hat{D}}}
\def\Gh{{\hat{G}}}
\def\Fh{{\hat{F}}}
\def\Ih{{\hat{I}}} 
\def\Jh{{\hat{J}}} 
\def\Kh{{\hat{K}}}
\def\Lh{{\hat{L}}} 
\def\Ph{{\hat{P}}}
\def\Rh{{\hat{R}}}
\def\Vh{{\hat{V}}} 
\def\Xh{{\hat{X}}}
 
\def\ah{{\hat{\a}}}
\def\bh{{\hat{\b}}}
\def\gh{{\hat{\g}}}
\def\dh{{\hat{\d}}}
\def\hh{\hat{h}}
\def\uh{\hat{u}}  
\def\xh{\hat{x}}  
\def\yh{\hat{y}}  
\def\ph{\hat{p}}  
\def\xih{\hat{\xi}}  
\def\chih{\hat{\chi}}  
\def\Psih{\hat{\Psi}}    
\def\phih{\hat{\phi}}

\def\psit{\tilde{\psi}}  
\def\Psit{\tilde{\Psi}}   
\def\Psibt{\tilde{\bar{Psi}}}  

\def\st{\tilde{\sigma}}  

\def\delt{\tilde{\delta}}
\def\Phit{\tilde{\Phi}}   
\def\Phitb{\overline{\tilde{Phi}}}  
\def\tht{\tilde{\th}}  
\def\lt{\tilde{\l}}
\def\chit{\tilde{\chi}}   
\def\phit{\tilde{\phi}} 

\def\At{\tilde{A}}
\def\Bt{\tilde{B}}
\def\Ct{\tilde{C}}
\def\Dt{\tilde{D}}
\def\Et{\tilde{E}}
\def\Ft{\tilde{F}}
\def\Gt{\tilde{G}}
\def\Ht{\tilde{H}}
\def\It{\tilde{I}}
\def\Jt{\tilde{J}}
\def\Qt{\tilde{Q}}  
\def\Rt{\tilde{R}}  
\def\Mt{\tilde{M }}  
\def\Nt{\tilde{N}}   
\def\St{\tilde{S}}
\def\Vt{\tilde{V}}
\def\Xt{\tilde{X}} 
\def\at{\tilde{a}}
\def\ct{\tilde{c}}
\def\dt{\tilde{d}}
\def\htt{\tilde{h}} 
\def\ft{\tilde{f}}
\def\gt{\tilde{g}}
\def\pt{\tilde{p}}  
\def\qt{\tilde{q}}  
\def\vt{\tilde{v}}  
\def\nt{\tilde{n}}  
\def\ut{\tilde{u}}  
\def\wt{\tilde{w}}  
\def\zt{\tilde{z}} 
\def\xt{\tilde{x}} 
\def\yt{\tilde{y}} 
\def\Psit{\tilde{\Psi}}
\def\vphit{\tilde{\varphi}}

\def\cHt{\tilde{\cH}}

\def\eb{\bar{\epsilon}} 
\def\delb{\bar{\partial}}  
\def\thb{\bar{\theta}}
\def\mub{\bar{\mu}}
\def\lamb{\bar{\l}}
\def\psib{\bar{\psi}}
\def\sb{\bar{\sigma}}
\def\xib{\bar{\xi}}
\def\chib{\bar{\chi}}

\def\Psib{\bar{\Psi}}
\def\Phib{\bar{\Phi}}
\def\Lamb{\bar{\Lambda}}
\def\Sb{{\overline \Sigma}}
\def\cb{\bar{c}}
\def\hb{\bar{h}}
\def\qb{\bar{q}}
\def\wb{\bar{w}}
\def\ub{\bar{u}}
\def\zb{{\bar{z}}}
\def\Hb{\bar{H}}
\def\Qb{{\bar Q}}
\def\Omegab{\overline{\Omega}}
\def\ob{\overline{\omega}}

\def\Ab{{\overline A}} \def\Bb{{\overline B}} \def\Cb{{\overline C}}  
\def\Db{{\overline D}} \def\Eb{{\overline E}} \def\Fb{{\overline F}}  
\def\Gb{{\overline G}} 
\def\Ib{{\overline I}}  
\def\Jb{{\overline J}} \def\Kb{{\overline K}} \def\Lb{{\overline L}}  
\def\Mb{{\overline M}} \def\Nb{{\overline N}} \def\Ob{{\overline O}}  
\def\Pb{{\overline P}}  \def\Rb{{\overline R}}  
 \def\Tb{{\overline T}} \def\Ub{{\overline U}}  
\def\Vb{{\overline V}} \def\Wb{{\overline W}} \def\Xb{{\overline X}}  
\def\Yb{{\overline Y}} \def\Zb{{\overline Z}}  

\def\fb{{\overline f}}
\def\gb{{\overline g}}
\def\mb{{\overline m}}
\def\lb{{\overline l}}
\def\yb{{\overline y}}
  
\def\ldel{{\overleftarrow{\del}}}
\def\rdel{{\overrightarrow{\del}}}
\def\ldeldel{{\overleftarrow{\del^2}}}
\def\rdeldel{{\overrightarrow{\del^2}}}
\def\ldelb{{\overleftarrow{\bar{\del}}}}
\def\rdelb{{\overrightarrow{\bar{\del}}}}

\def\ba{{\bf a}} 
\def\bk{{\bf k}}  
\def\bl{{\bf l}}  
\def\bp{{\bf p}}  
\def\bq{{\bf q}}  
\def\br{{\bf r}}
\def\bt{{\bf t}}
\def\bu{{\bf u}}
\def\bv{{\bf v}}
\def\bx{{\bf x}}  
\def\by{{\bf y}}  
\def\bR{{\bf R}}  
\def\bV{{\bf V}}

\def\bone{{\bf 1}}  

\def\va{{\vec a}}
\def\vk{{\vec k}}
\def\vp{{\vec p}}
\def\vq{{\vec q}}
\def\vx{{\vec x}}
\def\vy{{\vec y}}
\def\vu{{\vec u}}
\def\vv{{\vec v}}

\def\vs{{\vec \sigma}}
\def\vtau{{\vec \tau}}

\newcommand{\ov}[1]{\overrightarrow{#1}}

\def\frA{\mathfrak{A}}
\def\frB{\mathfrak{B}}
\def\frC{\mathfrak{C}}
\def\frD{\mathfrak{D}}
\def\frE{\mathfrak{E}}
\def\frF{\mathfrak{F}}
\def\frG{\mathfrak{G}}
\def\frH{\mathfrak{H}}
\def\frM{\mathfrak{M}}
\def\frN{\mathfrak{N}}
\def\frR{\mathfrak{R}}
\def\frW{\mathfrak{W}}

\def\fra{\mathfrak{a}}
\def\frb{\mathfrak{b}}
\def\frf{\mathfrak{f}}
\def\frg{\mathfrak{g}}
\def\frh{\mathfrak{h}}
\def\frl{\mathfrak{l}}
\def\frs{\mathfrak{s}}
\def\fri{\mathfrak{i}}
\def\frj{\mathfrak{j}}

\def\ma{\mathfrak{a}}
\def\mg{\mathfrak{g}}
\def\mh{\mathfrak{h}}
\def\mR{\mathfrak{R}}
\def\mN{\mathfrak{N}}

\def\d{\delta}\def\D{\Delta}\def\ddt{\dot\delta}  
  
\def\pa{\partial} \def\del{\partial}  
\def\xx{\times}  
\def\uno{\mbox{1 \kern-.59em {\rm l}}}    
  
\def\trp{^{\top}}  
\def\inv{^{-1}}  
\def\dag{{^{\dagger}}}  
\def\pr{^{\prime}}  
  
\def\rar{\rightarrow}  
\def\lar{\leftarrow}  
\def\lrar{\leftrightarrow}  
  
\newcommand{\0}{\,\!}      
\def\one{1\!\!1\,\,}  
\def\im{\imath}  
\def\jm{\jmath}  
  
\newcommand{\tr}{\mbox{tr}}  
\newcommand{\slsh}[1]{/ \!\!\!\! #1}  
  
\def\vac{|0\rangle}  
\def\lvac{\langle 0|}  
  
\def\hlf{\frac{1}{2}}  
\def\ove#1{\frac{1}{#1}}  

\def\Box{\square}  
\def\CC {\mathbb{C}}
\def\FF {\mathbb{F}}
\def\RR{\mathbb{R}}
\def\NN{\mathbb{N}}  
\def\ZZ{\mathbb{Z}}  
\def\bb#1{{\bf #1}}  
\def\bcomment#1{}  
\def\bfhat#1{{\bf \hat{#1}}}  
\def\VEV#1{\left\langle #1\right\rangle}  

\newcommand{\ex}[1]{{\rm e}^{#1}} \def\ii{{\rm i}}  

\newcommand{\lrbrk}[1]{\left(#1\right)}
\newcommand{\sfrac}[2]{{\textstyle\frac{#1}{#2}}}
 
\def\stw{{\sqrt{2}}}

\def\rf {{\rm f}}
\def\ri {{\rm i}}
\def\rj {{\rm j}}
\def\rk {{\rm k}}
\def\rl {{\rm l}}
\def\rs {{\scriptscriptstyle \rm S}}
\def\rt {{\scriptscriptstyle \rm T}}

\def\rQ {{\scriptscriptstyle \rm \cQ}}
\def\rR {{\scriptscriptstyle \rm \cR}}

\def\cBb{{\cal \Bb}}
\def\cQb{{\cal \Qb}}
\def\cRb{{\cal \Rb}}
\def\cWb{{\cal \Wb}}

\def\fd {{\rm N}}
\def\afd {{\overline{\rm N}}}

\def \II {I\hspace{-.1em}I\hspace{.1em}}
\def \IIA {\mbox{\II A\hspace{.2em}}}
\def \IIB {\mbox{\II B\hspace{.2em}}}
\def \gs {g^s}
\def \ls {\lambda^s}

\def \I {{\cal I}}
\def \qs {q\hspace{-.53em}/\hspace{.15em}}
\def \ks {k\hspace{-.53em}/\hspace{.15em}}
\def \YM {{\mbox{\tiny YM}}}
\def \gym {g_{\YM}}

\def \Lc {\L_c}
\def\IR{\relax{\rm I\kern-.18em R}}
\def \id {{\bf 1}}

\def\cci{\ell}
\def\ccj{\ell'}

\def \thbb{\overline{\th\th}}
\newcommand \ol{\overline}
\def \lamb{\bar{\lambda}}
\def \vphi{\varphi}
\def \lambh{\hat{\bar{\lambda}}}
\def \lh{\hat{\lambda}}
\def \dd{\ddagger}

\def \Xd{\dot{X}}
\def \nd{\noindent \\}

\author{Chong-Sun Chu, Sheng-Lan Ko 
 and Pichet Vanichchapongjaroen\\  
Centre for Particle Theory
and Department of Mathematical Sciences,\\ 
Durham University, Durham, DH1 3LE, UK \\
E-mail:  
\email{chong-sun.chu@durham.ac.uk}, \email{sheng-lan.ko@durham.ac.uk},
\email{pichet.vanichchapongjaroen@durham.ac.uk}}

\title{Non-Abelian Self-Dual String Solutions}

\abstract{
We consider the equations of motion of the 
non-abelian 5-branes theory recently constructed in 
\href{http://arxiv.org/abs/arXiv:1203.4224}{\cite{CK}}
and find exact string solutions both for 
uncompactified and compactified spacetime. 
Although one does not have the full 
supersymmetric construction of the non-abelian (2,0) theory, by
combining knowledge of conformal symmetry and R-symmetry  one can
argue for the form of the 1/2 BPS equations in the case when only one scalar
field is turned on. We solve this system and show that our string solutions 
could be lifted to become  solutions of the non-abelian (2,0) theory
with  self-dual electric and magnetic charges,
with the scalar field describing  a M2-brane spike emerging out of the 
multiple M5-branes worldvolume.
}
\preprint{DCPT-12/23}
\keywords{M-Theory, D-branes, M-branes, Gauge Symmetry}

\begin{document}
\section{Introduction}

The low energy theory of $N$ coincident M5-branes 
is given by an interacting (2,0) superconformal
theory in 6 dimensions \cite{zero}. 
On the M5-brane worldvolume there are self-dual strings.
For a single M5-brane, the low energy theory
is known \cite{howe,PS,schw1,pst,nilsson}. The self-dual string
soliton has also been constructed \cite{PS,HLW}.  Much less is known
about the theory of multiple M5-branes, as well as the properties of 
multiple self-dual strings. 
 
Recently, a theory of non-abelian chiral 2-form in 6-dimensions was
constructed \cite{CK}.
The construction was motivated by the 
analysis in \cite{chu,CS} and 
a set of 5d Yang-Mills gauge fields was introduced in order to
incorporate non-trivial interactions among the 2-form potential.  
The theory admits a self-duality equation on the field strength 
as the equation of motion. It has a modified 6d Lorentz symmetry. 
On dimensional reduction on a circle, the action gives the standard 5d 
Yang-Mills action plus higher order corrections.
Based on these properties, it was proposed that the theory describes 
the gauge sector of 
multiple M5-branes in flat space. An important feature of this theory
is that the self-interaction of the two-form gauge field is mediated
by a set of five-dimensional Yang-Mills gauge field $A_\mu, \mu
=0,1,2,3,4$). The Yang-Mills
gauge field is auxiliary and is constrained non-trivially to be 
given in terms of the non-abelian
tensor gauge field and does not contain any propagating 
degrees of freedom. In the Abelian case, the 1-form gauge field is
free and simply decouple. 
See also \cite{lam}, \cite{dou,lam2}, \cite{ho}, \cite{CG},
\cite{huang}, \cite{lee, sethi},
\cite{chu, sezgin, george, nishino},  
for some other more relevant recent developments. 

In this paper we give a further support of this proposal by
constructing the non-abelian self-dual strings  to
the equation of motion of the non-abelian theory \cite{CS}. 
Without loss of
generality, we consider a $SU(2)$ gauge group which corresponds to a
system of two M5-branes. A crucial observation in our construction is
that the Perry-Schwarz solution is supported by a Dirac monopole $A_a,
a=0,1,2,3$). As
the solution is translational invariant along the direction (say
$x^4$) of the string, this gauge field can be thought of as 
a five dimensional one with $A_4=0$ and be interpreted  as the
auxiliary 1-form gauge fields in the theory of \cite{CK}. 
This interpretation suggests that
the non-abelian self-dual string solution may be constructed by taking
the auxiliary Yang-Mills gauge field to be given by a non-abelian 
monopole. Quite remarkably this is indeed correct and we are able to 
construct a self-dual string solution both for uncompactified six
dimensions as well as with one dimension compactified. 
Our solution is
obtained by replacing the Dirac monopole in the 
Perry-Schwarz string, in the uncompactified case 
to the non-abelian Wu-Yang monopole; and in the compactified
case to the 't Hooft-Polyakov monopole.

The plan of the paper is as follows.
In section 2, we review the non-abelian 5-brane theory of \cite{CK}. 
In section 3, after reviewing the 
original Perry-Schwarz self-dual string
solution, we present a new abelian self-dual string solution which is
orientated in a different direction.   
The existence of the latter solution is guaranteed by the Lorentz
symmetry of the Perry-Schwarz theory. Then we solve the non-abelian 
equation of motion of \cite{CK} and obtain an exact solution describing
a string. We then discuss how this solution can be lifted as a solution
of the (2,0) supersymmetric theory. The resulting solution describes a
non-abelian string with self-dual charges. In section 4, we
consider the compactified case and construct the corresponding
self-dual string solution. The paper is concluded
with some further comments and discussions in section 5.

\section{Review of the Non-Abelian Multiple 5-brane Theory}

In \cite{CK}, an action for non-abelian chiral 2-form in 6-dimensions was 
constructed as a generalization of the linear theory of Perry-Schwarz. 
As in Perry-Schwarz,
manifest 6d Lorentz symmetry was given up and 
the self-dual tensor gauge field is represented 
by a $5 \times 5$ antisymmetric field $B_{\m\n}$, $\m, \n = 0,..,4$. 
Throughout the paper we use the convention that
the 5d and 6d coordinates are denoted by $x^\m= (x^0, x^1, \cdots, x^4)$ and
$x^M= (x^\m, x^5)$. We use $\eta^{MN} = (-+++++)$ 
for the metric and
$ \e^{01234} = - \e_{01234} =1$, $\quad \e^{012345} = - \e_{012345} =1$
for the antisymmetric tensors.
The Hodge dual of a 3-form $G_{MNP}$ is defined by
\be \label{hodge}
 \tilde{G}_{MNP} := - \frac{1}{6} \e_{MNPQRS}\, G^{QRS}.
\ee  

Motivated by the consideration in \cite{chu}, a set of 5d 1-form gauge fields
$A^a_\m$  was introduced for a gauge group $G$. The proposed action is 
\be
S = S_0 + S_E
\ee
with $S_0$ a non-abelian generalization of the Perry-Schwarz action, 
\be \label{S-PS-na}
S_0 = \frac{1}{2}\int d^6x\,\text{tr}\left( - \Ht^{\m\n}\Ht_{\m\n} 
+ \Ht^{\m\n}\del_5 B_{\m\n}\right)
\ee
where $H_{\m\n\l} = D_{[\m} B_{\n\l]} = [\del_{[\m} + A_{[\m}, B_{\n\l]} ] $; 
and with $S_E$ 
\be\label{SE}
S_E = \int d^5x\,\text{tr}\left( (F_{\m\n} - c \int dx_5\,\Ht_{\m\n})E^{\m\n}
\right),
\ee 
where $E_{\m\n}(x^\l)$ is a 5d auxiliary field, providing a 
constraint such that $A_\m$ carries no extra degrees of freedom.
Here $c$ is a constant and it was taken to be 1 in
\cite{CK}. 
Actually one can take any nonzero value of $c$ and this makes no change 
to all the symmetries discusses in 
\cite{CK}. 
The only modification is the relation of the Yang-Mills coupling 
to the compactification radius, $g^2_{YM} = \pi R c^2$. 
In the
following we will show how the value of $c$ is fixed by the
requirement of charge quantization of our self-dual string solution.

Besides the Yang-Mills gauge symmetry, 
\be
        \d A_\m = \del_\m\L + [A_\m,\L],  \quad        
\d B_{\m\n} = [B_{\m\n},\L], \quad        
\d E_{\m\n} = [E_{\m\n},\L]
\ee 
for arbitrary $\L = \L(x^\l)$,
the action has the tensor gauge symmetry
\be
        \d_T A_\m = 0,\quad
        \d_T B_{\m\n} = \S_{\m\n}, \quad
        \d_T E_{\m\n} = 0,
\ee
for $\S_{\m\n}(x^M)$ satisfying $D_{[\l}\S_{\m\n]}=0$.
This form of symmetry first appears in  \cite{ho}.
As demonstrated in \cite{CK}, the theory has  
manifest 5d Lorentz symmetry and a modified 6d Lorentz symmetry. 
To establish those symmetries of the action, 
we take the field configuration satisfying the boundary conditions: 
\be
        D_\l B_{\m\n}, ~\del_5 B_{\m\n} \rightarrow 0 \quad \text{as } 
|x^M| \rightarrow \infty
\ee

With an appropriate fixing of this tensor gauge symmetry, one can turn the 
equation of motion of $B_{\m\n}$ into a first order self-duality condition: 
\be \label{sd-na}
        \Ht_{\m\n} = \del_5 B_{\m\n}.
\ee
The gauge field is auxiliary and is determined by the equation:
\be \label{FB1}
F_{\m\n} = c \int dx_5\,\Ht_{\m\n}
\ee
This constraint was inspired from the analysis 
of the dimensional reduction, 
in which one gets multiple D4-branes plus higher derivative correction terms.  
Notice that, on mass-shell, the constraint \eq{FB1} simply says that 
$F_{\m\n}$ is given by the boundary values of $B_{\m\n}$  for the uncompactified 
case: 
\be \label{FB-bdy}
F_{\m\n} = c( B_{\m\n}(x_5 = \infty) - B_{\m\n}(x_5 = -\infty))
\ee
and
\be \label{FB-bdy2}
F_{\m\n} = 2 \pi R c \Ht_{\m\n}^{(0)},
\ee
when $x^5$ is compactified on a circle of radius $R$. Here $\Ht^{(0)}_{\m\n}$ 
is the zero mode part of the field strength.

\section{Non-Abelian Self-Dual String Solution: 
Uncompactified Case}

In this section, 
we construct  self-dual string  solution that satisfies both 
\eq{sd-na} and \eq{FB-bdy}. 
As mentioned above, a direct observation on the constraint \eq{FB-bdy}
shows that the solution cannot be aligned in the $x^5$ 
direction since this would imply
$F_{\m\n} =0$ which is trivial. 
This does not imply the non-existence 
of a string solution in other directions, 
because the self-duality equation \eq{sd-na} has only 5d Lorentz symmetry
as it's a gauge fixed equation of motion \cite{CK}.  
Therefore, as a preparation to constructing the more general
non-abelian self-dual string solution,  
we will first construct an abelian self-dual string solution
aligning in the $x^4$ direction and we will start by 
reviewing the original  abelian self-dual string solution of Perry and Schwarz.

\subsection{Self-dual string solution in the Perry-Schwarz Theory}

In \cite{PS},  a 
nonlinear theory of chiral 2-form gauge field which results in the 
Born-Infeld action for a $U(1)$ gauge field when reduced to 5 dimensions was 
constructed.  
The Perry-Schwarz non-linear field equation is given by
\be\label{nonlinear}
\Ht_{\m\n}=\frac{(1-y_1)H_{\m\n5}+H_{\m\r5}
H^{\r\s5}H_{\s\n5}}{\sqrt{1-y_1+\hlf y_1^2-y_2}},
\ee
where
\be
y_1:= -\hlf H_{\m\n 5}H^{\m\n 5}, \qquad
y_2:= \ove{4}H_{\m\n 5}H^{\n\r 5}H_{\r\s 5}H^{\s\m 5}.
\ee
As they demonstrated, the equation of motion \eq{nonlinear}  
admits a solution describing a self-dual string 
soliton with finite tension aligning in the direction $x^5$. Since  
\eq{nonlinear} is (non-manifest) 6d Lorentz covariant,
it means there must also exist 
self-dual string solution aligned in other 
directions. In the following, we review their construction in
section \ref{soln-x5}. Then we construct new self-dual string 
solution aligned in 
a different direction in section \ref{soln-x4}.

\subsubsection{Self-dual string in the $x^5$ direction}
\label{soln-x5}

The ansatz Perry and Schwarz considered for their self-dual string solution is
\be \label{B-PS}
B=\a(\r) dt dx^5+\frac{\b}{8}(\pm 1-\cos\tht)d\phit d\psit,
\ee
where the 6d metric is 
\be
ds^2=-dt^2+(dx^5)^2+d\r^2+\r^2d\Omega_3^2,
\ee
with the three-sphere given in Euler coordinates
\be \label{g-Euler}
d\Omega_3^2=\ove{4}[(d\psit+\cos\tht d\phit)^2+(d\tht^2+\sin^2\tht d\phit^2)],
\ee
where $0\leq\tht\leq\pi, 0\leq\phit\leq 2\pi, 0\leq\psit\leq 4\pi.$
For this ansatz, it is $y_1 = \a'\,{}^2$, $y_2 = \a'\,{}^4/2$
and the non-linear field equation \eq{nonlinear} reads
\be \label{a1}
\a'(\r)=\frac{\b}{\sqrt{\b^2+\r^6}}.
\ee
This can be solved easily in terms of a hyper-geometric function. 
The solution is regular everywhere where $\a \sim \r$ as $\r \to 0$, 
while $\a \sim -\frac{\b}{2 \r^2} + {\rm const.}$ as $\r \to \infty$.
Note that the same ansatz also solves
the linear self-duality equation, where
in this case we have,
\be \label{a2}
\a'(\r)=\frac{\b}{\r^3}
\ee
and the solution is singular at $\r =0$. 
In other words,
the non-linear terms in the field equation has smoothen out the 
singularity at $\r=0$. 

The  magnetic charge $P$ and 
electric charge $Q$ per unit length of the string are given by
\be \label{PQ-def}
P=\int_{S^3} H, \qquad Q=\int_{S^3} *H, 
\ee
where $*$ denotes the Hodge dual operation and $S^3$ is a three sphere 
surrounding the string. It is straightforward
to obtain that  
\be\label{PQ}
P = 2\pi^2\b, \quad \mbox{and}\quad
Q = 2\pi^2\r^3\a'(\r)|_{\r\to\infty}=2\pi^2\b,
\ee 
hence the string is self-dual.
This holds for both
the nonlinear and the linear cases. 
Note that our answer is $1/8$ of those in \cite{PS} as  
we have introduced the factor of $1/4$ into the metric \eq{g-Euler} in order to 
reproduce the correct volume $2\pi^2$ for a unit three sphere.

The charge quantization condition
\cite{Deser:1997mz,Deser:1997se}
\be \label{PQ-quan}
PQ+QP \in 2\pi {\bf Z}
\ee
for the self-dual string gives
\be
\b=\pm\sqrt{\frac{n}{4\pi^3}},
\ee
i.e.
\be
P=Q = \pm \sqrt{n \pi}, 
\ee
where $n$ is a positive integer.
Note that the charge quantization condition we used is
different from the Dirac-Teitelboim-Nepomechie
charge quantization condition \cite{d1,d2,d3}
Perry and Schwarz used. 
The condition \eq{PQ-quan} is obtained with a self-dual string probing 
another self-dual string and 
the positive sign in the charge quantization condition
is appropriate for dyonic branes in 
$D=4k+2$ spacetime dimensions \cite{Deser:1997mz,Deser:1997se}.

Perry and Schwarz have also computed the tension
of their string solution. Since the solution is static, the energy can be 
identified with the Lagrangian and the energy per unit length is found to be
\be
T  = \ct \b^{4/3}, 
\ee 
where $\ct$ is a numerical coefficient. 
We remark that for the self-dual string solution of the linearized theory,
the tension is 
\be
T =0
\ee
since obviously the action
vanishes on-shell. Since the charges and tension are well defined, 
it appears that the singularity at $\rho=0$ is not harmful.

We also remark that 
the Perry-Schwarz self-dual string solution is non-BPS as there is no 
other matter field turned on to cancel the tensor field force. 
In the literature,
there is also the 1/2 BPS self-dual string of Howe, Lambert and West 
\cite{HLW}.
In fact the Perry-Schwarz self-duality equation of motion can be
embedded in the fully supersymmetric five-brane equation of motion of 
\cite{howe} by setting all the matter fields to zero and hence 
the Perry-Schwarz
self-dual string solution can be lifted to be a solution of the full 
five-brane
equation of motion, albeit a nonsupersymmetric one. Unlike the 
nonlinear Perry-Schwarz self-dual string solution, 
the Howe-Lambert-West self-dual string solution is singular at the location of 
the string. In fact $B \sim 1/\r^2$ near the string, which is exactly as in 
linearized Perry-Schwarz self-dual
string solution.

\subsubsection{Self-dual string soliton in the $x^4$ direction}
\label{soln-x4}

The Perry-Schwarz solution is translationally invariant along $x^5$.
One may want to generalize this solution directly and construct a 
non-Abelian self-dual string solution which is 
translationally invariant along $x^5$ but this is not possible.
As reviewed above, the gauge field strength 
in the non-abelian theory
is given on-shell by
the boundary value of $B$-field as \eq{FB-bdy}
Therefore, if the non-Abelian solution is translationally 
invariant along $x^5$,
then $F_{\m\n}=0$ which is trivial. 

To get a non-trivial solution, we 
need to base
our construction on Perry-Schwarz solitons which are translationally invariant
along other direction, say $x^4$. 
Such a solution can be easily obtained by rotating the original Perry-Schwarz 
solution as Perry and Schwarz has proved that their theory and 
the non-linear equation \eq{nonlinear} respect
Lorentz symmetry. Therefore, a simple Lorentz transformation 
which swap $(x_4,x_5)\rightarrow(-x_5,x_4)$
can be applied
on the original Perry-Schwarz solution 
(the minus sign is needed to preserve the orientation of spacetime) 
to obtain
the desired solution.

To facilitate the discussion, it is more convenient to use the spherical polar 
coordinates
which is related to the Euler coordinates by the change of coordinates
\be
\tht=2\th,\qquad \phit=\psi-\f,\qquad \psit=\psi+\f.
\ee
With this coordinates, the three-sphere metric is given by
\be
d\Omega_3^2=d\th^2+\sin^2\th d\f^2+\cos^2\th d\psi^2
\ee
with the ranges $0\leq\th\leq\pi/2, 0\leq \f,\psi\leq 2\pi$,
and the Perry-Schwarz ansatz \eq{B-PS} becomes
\be
B=\a(\r) dt dx^5+\b\lrbrk{\ove{4}\pm \ove{4}-\hlf\cos^2\th}d\phi d\psi.
\ee
Next change to Cartesian coordinates 
\be
x=\r\sin\th\cos\f, \quad
y=\r\sin\th\sin\f, \quad
z=\r\cos\th\cos\psi,\quad
w=\r\cos\th\sin\psi,
\ee
where we have denoted $(x^1,x^2,x^3,x^4)=(x,y,z,w)$. 
The metric becomes
\be
ds^2=-dt^2+dx^2+dy^2+dz^2+dw^2+d(x^5)^2,
\ee
and the Perry-Schwarz ansatz reads
\be \label{B-PS1}
B=\a(\r)dt dx^5+\b\frac{\ove{4}\pm\ove{4}-\frac{1}{2}
\frac{w^2+z^2}{\r^2}}{(x^2+y^2)(z^2+w^2)}(xzdydw-xwdydz-yzdxdw+ywdxdz).
\ee
 
Keeping the orientation, we swap 
$(x_4,x_5) \rightarrow (-x_5,x_4)$
and obtain our ansatz for a string solution along
the $x^4$ direction,
\be \label{B-PS2}
B=\a(\r)dt dw - 
\b\frac{\ove{4}\pm\ove{4}-\frac{1}{2}
\frac{(x^5)^2+z^2}{\r^2}}
{(x^2+y^2)(z^2+(x^5)^2)}(xzdydx^5-xx^5dydz-yzdxdx^5+yx^5dxdz)
\ee
where now
\be
\r=\sqrt{(x^5)^2+r^2}, \quad r:= \sqrt{x^2+y^2+z^2}.
\ee
It follows that
\be \label{H-ab}
H=\frac{\a'}{\r}dt dw\big(xdx+ydy+zdz+x^5dx^5\big)+
\frac{\b}{\r^4}\big(x^5dxdydz-xdydzdx^5+zdydxdx^5-ydzdxdx^5\big),
\ee
\be
*H=\frac{\a'}{\r}\big(x^5dxdydz-xdydzdx^5+zdydxdx^5-ydzdxdx^5\big)+
\frac{\b}{\r^4}dt dw\big(xdx+ydy+zdz+x^5dx^5\big),
\ee 
and
\be
y_1=\frac{(\a')^2(x^5)^2}{\r^2}-\frac{\b^2 r^2 }{\r^8}, \qquad
y_2=\frac{\b^4 r^4}{2\r^{16}}+\frac{(\a')^4(x^5)^4}{2 \r^4}.
\ee
Then the field equation \eq{nonlinear} gives
\be \label{nl1}
\frac{\b}{\r^4}x^5dt dw+\frac{\a'}{\r}
(-xdydz+zdydx-ydzdx)=\frac{\a'x^5}{\r}Gdt dw+
\frac{1}{G}\frac{\b}{\r^4}(-xdydz+zdydx-ydzdx),
\ee
where
\be
G=\sqrt{\frac{1+ \b^2 r^2 \r^{-8}}{1-\a'^2(x^5)^2 \r^{-2}}}.
\ee
The equation \eq{nl1} is equivalent to
\be
\a'=\frac{\b}{\sqrt{\b^2+\r^6}},
\ee
which is the same equation as before.
As a consistency check, we integrate over the $S^3$ transverses to $x^4$
and obtain the same charges
\be
P=Q=2\pi^2\b.
\ee For the
linearized case, $\a' = \b/\r^3$. 

\subsubsection{Self-dual string soliton in the $x^4$ direction in 
the $B_{\m 5}=0$ gauge}

The potential  $B_{MN}$ in the solution \eq{B-PS1} or \eq{B-PS2}
does not satisfy the condition $B_{\m 5}=0$ as needed in \cite{PS,CK}.
However this is not a problem as they are indeed gauge equivalent to one which 
does.
Instead of giving the gauge transformation, it is more instructive to
construct directly
the linearized self-dual string soliton in the $x^4$ direction in this gauge.

The starting point is
\eq{H-ab} with $\a' = \b/\r^3.$ Our strategy is to integrate 
the self-duality equation of motion
\be \label{sd-eom0}
H_{\m\n 5}=\pa_5B_{\m\n}
\ee
to get $B_{\m\n}$. Then we use $B_{\m\n}$ to compute the whole $H_{MNP}$ and
check its consistency with our ansatz. 
The components of $H$ are
\be\label{H1-ab}
H_{twi}=\frac{\b x^i}{\r^4}, \qquad
H_{ijk}=\frac{\e_{ijk}\b x^5}{\r^4},
\ee
\be \label{H2-ab}
H_{tw5}=\frac{\b x^5}{\r^4}, \qquad
H_{ij5}=-\frac{\e_{ijk}\b x^k}{\r^4}.
\ee
Integrating \eq{H2-ab}, we get the following
components of $B_{\m\n}:$
\be \label{soln-B1-ab}
B_{ij}=-\hlf\frac{\b\e_{ijk}x_k}{r^3}\lrbrk{\frac{x^5r}{\r^2}
+\tan^{-1}(x^5/r)}, \qquad
B_{tw}=-\frac{\b}{2\r^2},
\ee
In principle, $x^5$ independent constants of integration can be added 
but we will not need them. 
It is now easy to check a consistent solution is
obtained by setting all the other independent components 
of $B_{\m\n}$ to be zero. 

Two remarks are in order:
\bit
\item[1.]
We remark that if we apply  
the condition \eq{FB1} to the Perry-Schwarz self-dual string solution, we obtain
\be
F_{ij}=-\frac{c\b\pi}{2}\frac{\e_{ijk}x_k}{r^3},
\qquad F_{tw}=0
\ee 
for the auxiliary gauge field. 
Certainly this $U(1)$ field decouples and play no role in the abelian case. 
However it is interesting to note that 
this is  precisely the field strength of a Dirac monopole 
in the $(x,y,z)$ subspace! 
The presence of a Dirac monopole was already apparent in the
original solution of \cite{PS}. 
Here, we reveal that the same monopole configuration 
also appears as the auxiliary gauge field.
It turns out the use of an non-abelian monopole in place of the Dirac monopole
is precisely what is needed to construct the non-abelian self-dual string solution.

\item[2.]
The solution in the form \eq{soln-B1-ab} 
will be our basis for the construction of
the non-abelian self-dual string in the next subsection.
We remark that it is also quite interesting 
that this form of the solution
provides a link
between linearized Perry-Schwarz self-dual string
and Howe-Lambert-West self-dual string \cite{HLW}. To explain this,
let us first give a brief review on the key construction
of Howe-Lambert-West self-dual string.
In the (2,0) supersymmetric theory, there are
two non-linearly related 3-form field strengths
which are called $H$ and $h.$ The 3-form $H$ is
exact but not necessarily self-dual while the
3-form $h$ is self-dual but not necessarily
exact. When constructing self-dual string,
one of the scalar fields is also turned on. 
The equation of motion is non-linear.
However, with an appropriate ansatz, it
is possible to impose a BPS condition
which eventually gives a linear differential 
relation between $H$ and the scalar field.
Writing in our notation, the BPS equations of motion read
\be\label{ab-BPS1}
H_{twi}=\pa_i\f,\qquad
H_{tw5}=\pa_5\f,
\ee
\be\label{ab-BPS2}
H_{ijk}=\e_{ijk}\pa_5\f,\qquad
H_{ij5}=-\e_{ijk}\pa_k\f,
\ee
where we have rescaled the scalar to absorb an inessential numerical factor.
These conditions ensure the self-duality of  $H$. Furthermore, they agree
precisely with the Perry-Schwarz's equations of motion \eq{sd-eom0} if
one identifies $B_{tw}=\f$.
In other words, the linearized Perry-Schwarz
self-dual string solution could be lifted to a
1/2 BPS solution in the (2,0) supersymmetric theory 
by adding a scalar field
that satisfies the \lq BPS\rq\ condition \eq{ab-BPS2}
(due to self-duality, the condition \eq{ab-BPS1} is
not needed).

\eit

\subsection{Non-abelian Wu-Yang string solution}
\label{soln-nab}

Now we are ready for the non-abelian case.
As noted above of the roles played by the Dirac monopole in the 
abelian Perry-Schwarz solution, it is natural to  consider the 
non-abelian generalizations of the Dirac monopole in the construction of the
non-abelian self-dual strings. Here we have two candidates:
the Wu-Yang monopole and the 't Hooft-Polyakov monopole where the latter
involves a Higgs scalar field while the former does not.
See, for example, 
\cite{shnir} for a review of these solutions. 
We will use these non-abelian configurations to
construct non-abelian self-dual string solutions for both the
uncompactified case (where the Wu-Yang solution will be used) 
and compactified case
(where the 't Hooft-Polyakov monopole will be used).

Let us first briefly review the non-abelian Wu-Yang monopole. 
Without loss of
generality, we will consider $SU(2)$ gauge group
with Hermitian generators $T^a = \frac{\s^a}{2}$ satisfying
\be
[T^a,T^b] = i \e^{abc} T^c, \quad a,b,c =1,2,3.
\ee
This corresponds to the relative gauge symmetry of a system of two five-branes. 
Our convention for the Lie algebra valued fields  are: 
$F_{\m\n}= i F_{\m\n}^a T^a$, $A_\m = i A_\m^a T^a$ and
$F_{\m\n}^a = \del_\m A_\n^a - \del_\n A_\m^a  - \e^{abc} A^b_\m A^c_\n$.

The non-abelian Wu-Yang monopole is given by 
\be \label{wy}
A_i^a= - \e_{aik}\frac{x_k}{r^2}, \qquad 
F_{ij}^a=\e_{ijm} \frac{x_mx_a}{r^4}, 
\ee
where $i,j =1,2,3$ and 
Note that the  field strength 
for the Wu-Yang solution is related to  the field strength 
$F_{ij}^{\rm (Dirac)} =  \e_{ijm}x_m / r^3$ 
of the Dirac monopole by a simple relation:
\be \label{wu-yang-F}
F^a_{ij}=    F_{ij}^{\rm (Dirac)}  \frac{x^a}{r}.
\ee
In fact by
performing a (singular) gauge transformation
\be \label{sing-U}
U= e^{i \s_3 \vphi/2} e^{ i\s_2 \th/2} e^{-i \s_3 \vphi/2},
\ee
one can go to an Abelian gauge where only the 3rd component of the gauge
field survives. 
In this gauge
\be \label{AA}
A_i^a = \d^a_3 \; A_i^{(\rm Dirac )} .
\ee
Despite its close connection with the Dirac monopole, 
the Wu-Yang solution is not a monopole since it does not 
source the non-abelian magnetic field. In fact the color magnetic
charge vanishes 
\be \label{chern}
\int_{S^2} F^a =0.
\ee
Nevertheless the Wu-Yang solution 
is a useful prototype for constructing a non-abelian
monopole and we will follow the common practice of the literature to refer to 
it as the Wu-Yang monopole. In particular, a magnetic charge can be defined if 
there is also in presence a Higgs scalar field 
as in the 't Hooft-Polyakov monopole.

Inspired by the relation \eq{wu-yang-F} of the Wu-Yang solution, we will try to
solve the non-abelian self-duality equation \eq{sd-na} by adopting the 
following ansatz for the field strength,
\be\label{nab-H-ansatz}
H^a_{\m\n \l} =   H_{\m\n\l}^{(\rm PS)} \frac{x^a}{r} 
\ee
Here $r = \sqrt{x^2+y^2+z^2}$ and 
\bea
H^{\rm (PS)}:=\frac{\b}{\r^4} 
\Big[
dt dw(xdx &+& ydy+zdz+x^5dx^5) \nn\\
&+& x^5dxdydz-zdxdydx^5-ydzdxdx^5-xdydzdx^5
\Big] 
\eea
is the field strength for the linearized Perry-Schwarz solution 
in the $x^4$ direction \eq{H-ab}.
The self-duality of \eq{nab-H-ansatz} follows immediately from the self-duality 
of the Perry-Schwarz solution. For the moment, we will allow $\b$ to be 
a free parameter.

Our strategy is again to integrate  $H_{\m\n 5}=\pa_5B_{\m\n}$ 
to get $B_{\m\n}$. Then
we obtain $F_{\m\n}$ and $A_\m$ from the boundary value of $B_{\m\n}$.
Finally, we use $B_{\m\n}$ and $A_\m$ to compute the whole $H_{MNP}$ and
check its consistency with our ansatz.  
Now the components of our ansatz are:
\be\label{H1}
H^a_{twi}= \frac{\b x^ix^a}{r\r^4}, \qquad
H^a_{ijk}= \frac{\e_{ijk}\b x^5x^a}{r\r^4},
\ee
\be \label{H2}
H^a_{tw5}= \frac{\b x^5x^a}{r\r^4}, \qquad
H^a_{ij5}=-\frac{\e_{ijk}\b x^kx^a}{r\r^4}.
\ee
Integrating \eq{H2}, we get the following
components of $B_{\m\n}:$
\be \label{nab-B-PS}
B^a_{\m\n}=  
B_{\m\n}^{\rm (PS)}\frac{x^a}{r}, \quad \mbox{$\m\n = ij$ or $tw$}, 
\ee
where $B^{(\rm PS)}_{ij}, B_{tw}^{(\rm PS)}$ 
are the $B$-field components \eq{soln-B1-ab} for the Perry-Schwarz solution. 
In principle, $x^5$ independent constants of integration can be added 
but we will not need them. 

A consistent solution can
be obtained by setting all the other independent components 
of $B_{\m\n}$ to be zero. 
To see this, let us compute $F_{\m\n}$ from \eq{FB-bdy}. It is remarkable that
\be
F^a_{ij}=  - \frac{c\b\pi}{2}\frac{\e_{ijm}x_m x_a }{r^4},
\qquad F^a_{tw}=0,
\ee 
which  is precisely the form \eq{wy} 
of the Wu-Yang monopole if we take
\be\label{WY-fix}
c\b = - \frac{2}{\pi}.
\ee
As a result, the non-vanishing component of the 
gauge field is given by 
\be \label{soln-A}
A_i^a= - \e_{aik}\frac{x_k}{r^2}.
\ee
So far we have used only  the field strength components $H_{ij5},
H_{tw5}$ of \eq{H2}. 
However since $D_\m(x^aT^a/r) = 0$ for the Wu-Yang gauge field, 
therefore \eq{H1} is reproduced immediately and
\eq{nab-H-ansatz} is indeed satisfied.

Like the Wu-Yang monopole, the color magnetic charge of our
Wu-Yang string solution vanishes.
This is not a problem as we should not forget about the scalar fields as
our ultimate aim is to construct the non-abelian self-dual string solution in
the multiple M5-branes theory and so the inclusion of 
scalar fields is natural from the point of view of
(2,0) supersymmetry. Although we do not have
the full (2,0) supersymmetric theory, one can argue that the self-duality
equation of motion \eq{sd-na} is not modified by the presence of the 
scalar fields. This can be seen by a simple dimensional analysis since 
the dimension of a canonically normalized scalar 
field is two, and there is no local polynomial term  one can write down 
which is consistent with conformal symmetry. That the self-duality
equation is not modified by the scalar fields is also the case in the
other proposed constructions \cite{lam2,sezgin}. As for the scalar field, 
first it is clear that due to R-symmetry, 
the self-interacting potential vanishes if there is
only one scalar field turned on. As a result, the equation of motion of the
scalar field is 
\be \label{eom-s}
D_M^2 \phi =0. 
\ee 
This is the general situation but for special cases, for example
when a BPS condition is satisfied, the 
second order equation could be reduced to a first order equation. 
A reasonable form of the BPS equation is the non-abelian generalization of
the BPS equation \eq{ab-BPS1}, \eq{ab-BPS2}
\be \label{bps}
H_{ijk}=\e_{ijk}\pa_5\f,\qquad
H_{ij5}=-\e_{ijk}D_k\f.
\ee 
We conjecture that \eq{bps} is indeed a BPS equation of the non-abelian 
(2,0) theory since first of all it   implies the equation of motion \eq{eom-s}. 
Moreover \eq{bps} would
follow immediately from the supersymmetry transformation 
($\G^{012345} \e =\e$, $\G^{012345} \psi = -\psi$)
\be
\d \psi = (\G^M \G^I D_M \phi^I + \frac{1}{3!2} \G^{MNP} H_{MNP}) \e
\ee
(which is the most natural non-abelian 
generalization of the abelian (2,0) supersymmetry transformation)
and the 1/2 BPS condition
\be
\G^{046} \e =-\e,
\ee
together with the condition that $\phi^6:=\phi =\phi (x^a), a=1,2,3,5$. 

\FIGURE[r]{
\epsfig{file=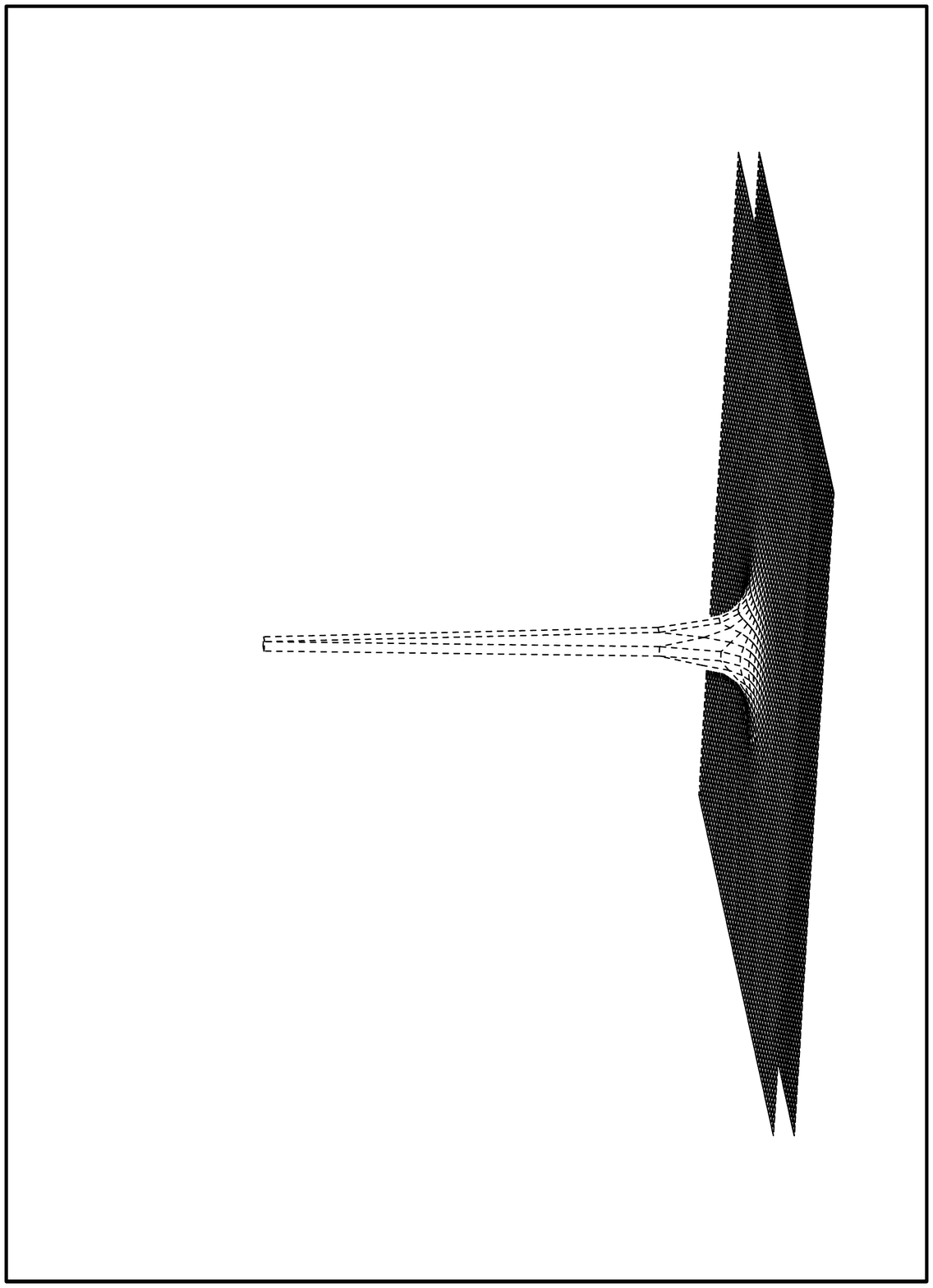,height=8.3cm}
\caption{An M2 brane ending on a system of two 
parallel M5-branes separated by a distance.}
\label{m5v}
}

We note that \eq{bps} is compatible with the self-duality equation 
if the scalar field is equal to the $B_{tw}$ component:
\be
\phi^a =B^a_{tw}= -\frac{\b}{2\r^2}\frac{x^a}{r},
\ee
or more generally,
\be\label{nab-s}
\phi^a=  -\lrbrk{u +\frac{\b}{2\r^2}}\frac{x^a}{r},
\ee
where $u$ is a constant. 
To see the physical meaning of this solution, let us
consider the transverse distance $|\phi|$ defined by $|\phi|^2=\phi^a \phi^a$.
This gives
\be
|\phi|= |u+\frac{\b}{2\r^2}|.
\ee
We will choose the constant $u$
to be of the same sign as $\b$ so that $|\phi|$ is never zero.
This describes 
a system of M5-branes with a spike at $\r=0$
and level off to $u$ as $\r\to\infty.$ Hence the
physical interpretation of our self-dual string is that
two M5-branes are separating by a distance $u$
and with an M2-brane ending on them. 
With this interpretation, there is a symmetry breaking
and one can identify an $U(1)$ $B$-field
at the large distance $\rho$:
\be
\cB_{\m\n}\equiv \phih^a B^a_{\m\n} = \pm B_{\m\n}^{(\rm PS)}
\ee
where $\phih^a := \phi^a/|\phi|$ and 
the $+$ $(-)$ sign in the second equation above corresponds to
the case $c>0$ $(c<0)$.
Since the field configuration approaches that of the abelian self-dual string 
at large distance, we immediately obtain the charges
\be
P= Q = - 2\pi^2 |\b| = -\frac{4 \pi}{|c|}.
\ee
and charge quantization determines that 
\be 
\b= \mp \sqrt{\frac{n}{4\pi^3}}, \quad
c= \pm 4\sqrt{\frac{\pi}{n}}
\ee
and $P=Q=  -\sqrt{n \pi}$.
We require that the theory should admit solution with the minimal 
unit of charge 
and so the 
possible values of the constant $c$ in the non-abelian action \eq{SE} is:
\be \label{constant-c}
c= \pm 4\sqrt{\pi}
\ee
and the charges of our solution are $P=Q= -\sqrt{\pi}$. 

Just as in the abelian case, the action for the gauge fields vanish on shell. 
Therefore the string gets its tension solely from the scalar field.  
In general, the kinetic term of scalar field
is proportional to
\be
\tr (D_M\f D^M\f).
\ee
Since the scalar field satisfies
\be
D_M \f\to 0,\qquad\r\to\infty,
\ee
we see that at large distance $\rho \to \infty$  from the string,
the kinetic term vanishes. However the singularity at the origin leads to
an infinite tension.
This is the same as the Howe-Lambert-West self-dual string solution 
\cite{HLW}.

\section{Non-Abelian Self-Dual String Solution: Compactified Case}

In this section, we consider the theory  
with $x^5$ compactified on a circle with radius $R$ and construct the 
self-dual string solution. The constraint that the gauge field has to satisfy
is now \eq{FB-bdy2}. Without loss of generality, let us assume that the string 
aligns in the $w=x^4$ direction.

In the compactified theory, 
the field strength can be expanded in terms of Fourier modes, 
\be
H_{MNP}=\sum_{n}e^{inx^5/R}H_{MNP}^{(n)}(r).
\ee
The gauge field $B_{\m\n}$ can be then obtained by integrating over the equation
of motion
$H_{\m\n5} = \del_5B_{\m\n}$. It is 
\be
B_{\m\n}=\frac{x^5}{2\pi Rc}F_{\m\n}(r)+\sum_{n = -\infty}^\infty
e^{inx^5/R}B_{\m\n}^{(n)}(r),
\ee
where 
we have used the boundary condition \eq{FB-bdy2} to determine the first term and
$B^{(0)}_{\m\n}(r)$ is an integration constant.
The higher modes $B^{(n \neq 0)}_{\m\n}$ are given by:
\be
H^{(n\neq 0)}_{\m\n 5}(r)=\frac{in}{R}B^{(n\neq 0)}_{\m\n}(r).
\ee
Notice that the first term on the right hand side has no contribution to 
$H_{\m\n\l}$ because of Bianchi identity and hence
\be
H_{\m\n\l}^{(n)}=D_{[\l}B_{\m\n]}^{(n)}
\ee
for all $n$. 

Let us consider an ansatz with the only nonzero components of gauge potential being 
$B_{tw}$ and $B_{ij}$. The self-duality condition reads 
\be
H_{ijk}=\e_{ijk}H_{tw5},
\quad H_{twk}=-\hlf\e_{ijk}H_{ij5},
\ee
or, written in terms of modes,
\be\label{sdzero}
D_{[i}B_{jk]}^{(0)}=\e_{ijk} \frac{F_{tw}}{2\pi R c},
\qquad 
D_kB_{tw}^{(0)}=-\frac{f_k}{2\pi R c} 
\ee
\be\label{sdkk}
D_kb_k^{(n)} =\frac{in}{R}B_{tw}^{(n)}, \quad   
\quad
D_kB_{tw}^{(n)} =-b_{k}^{(n)} \frac{in}{R}, \qquad n\neq 0,
\ee
where we have denoted 
\be
f_k(r) :=\hlf\e_{ijk}F_{ij} 
\qquad \mbox{and}\qquad 
b_k^{(n)}(r) :=\hlf\e_{ijk}B_{ij}^{(n)}\quad \mbox{for $n \neq 0$}.
\ee
Notice that the 2nd equation of (\ref{sdzero}) takes  exactly the
same form as the  BPS equation 
for the 't Hooft-Polyakov magnetic monopole 
if we identify $-2\pi R c B_{tw}^{(0)}$ 
as the scalar field there. 
Indeed in the BPS limit, 
the equation of motion for the 't Hooft-Polyakov monopole reads
\be
\hlf\e_{ijk}F_{ij}=D_k\phi,
\ee
where $\phi$ is an adjoint Higgs scalar field. 
The solution is given by
\be
A^a_i= - \e_{aik}\frac{x^k}{r^2}(1-k_v(r)),
\qquad 
\phi^a=\frac{vx^a}{r}h_v(r),
\ee
where
\be
k_v(r) :=\frac{vr}{\sinh(vr)},
\qquad 
h_v(r) := \coth(vr) - \frac{1}{vr}.
\ee
Asymptotically $r\to\infty,$ we have
\be
A^a_i\to - \e_{aik}\frac{x^k}{r^2},
\qquad 
\phi^a \to\frac{|v|x^a}{r}: = \phi_\infty,
\ee
which coincides with Wu-Yang monopole.
Note that the gauge symmetry is broken at infinity to $U(1)$, the little group
of $\phi_\infty$. This may be identified as the electromagnetic gauge group and 
one could use
this to define the magnetic monopole charge \cite{hooft,polyakov}.
The electromagnetic field strength can be defined as
\be
\cF_{ij}= F_{ij}^a \frac{\phi^a}{|v|}
=\e_{ijk}\frac{x^k}{r^3},\qquad \text{for large $r$}.
\ee
The magnetic charge is given by
$p=\int_{S^2}\cF=4\pi$,
which corresponds to a magnetic monopole of unit charge.
Note that at the core $r\to 0$, we have
\be
A_i\to 0,
\qquad \f\to 0
\ee
and hence the $SU(2)$ symmetry is unbroken at the monopole core.

The resemblance  of  our equation with the BPS equation of the 
't Hooft-Polyakov monopole 
motivates us to take for $A_\m$ the same ansatz  as in the 't Hooft-Polyakov 
monopole,
\be
A^a_i=-\e_{aik}\frac{x^k}{r^2}(1-k_v(r)),
\ee
This implies $F_{tw} = 0$ and hence the 1st equation of (\ref{sdzero}) 
can be solved with 
\be
B_{ij}^{(0)} = c_0 F_{ij},
\ee
where $c_0$ is an arbitrary constant.
On the other hand, (\ref{sdkk}) gives 
\be
D_kD_kB_{tw}^{(n\neq 0)}=\frac{n^2}{R^2}B_{tw}^{(n\neq 0)}.
\ee
For zero mode, 
we have $D_kD_k B_{tw}^{(0)} = 0$, combine them together we can write 
\be\label{comscalar}
D_kD_kB_{tw}^{(n)}=\frac{n^2}{R^2}B_{tw}^{(n)}.
\ee
We take the ansatz for $B_{tw}^{(n)}$ as 
\be 
B_{tw}^{(n) \;a}=a_n(r)\frac{vx^a }{r}
\ee
then the equation (\ref{comscalar}) is equivalent to
\be
\frac{\pa_r(r^2\pa_ra_n(r))}{r^2}-\frac{2k_v(r)^2}{r^2}a_n(r)=
\frac{n^2}{R^2}a_n(r).
\ee
The well-behaved physical solution is 
\be
a_0 = \a_0 h_v(r),
\ee
\be
a_{n\neq 0}(r)=\a_n\frac{e^{-|n|r/R}}{vr}\lrbrk{1+\frac{vR}{|n|}\coth (vr)},
\ee
where $\a_n$ are arbitrary constants. 
Here we have dropped the independent solutions which are exponentially 
increasing at large distance and hence not physical.
As a result, we obtain for the gauge fields
\be
B^a_{tw}=-\frac{h_v(r)}{2\pi R c}\frac{vx^a }{r}+\sum_{n\neq 0}
\a_ne^{inx^5/R}\frac{e^{-|n|r/R}}{vr}\lrbrk{1+\frac{vR}{|n|}\coth (vr)}
\frac{vx^a }{r},
\ee
\be
B^a_{ij}=\frac{x^5}{2\pi R c}F^a_{ij}(r)+c_0 F^a_{ij}(r)+\sum_{n\neq 0}
e^{inx^5/R}B_{ij}^{a\; (n)}(r).
\ee
where 
\be
b_{k}^{(n)\;a}=-v^3\frac{R}{in}(ra_n'-k_v(r)a_n)
\frac{x^kx^a }{r}-\d^a_k \frac{v R}{in }a_nk_v(r)\frac{1}{r}, \quad
n\neq 0.
\ee
The proportionality factor for $a_0$ is determined by recalling that 
$-2\pi RcB_{tw}^{(0)}$ is the scalar of the 't Hooft-Polyakov monopole, 
while $\a_{n\neq0}$ are left undetermined. 
Physically this corresponds to different excitations over the 
fundamental solution 
with all $\a_{n \neq 0} =0$. 
Note that there is a ``winding mode'' in $B_{ij}$, 
while there is no such mode in $B_{tw}$ because $F_{tw} = 0$. 
Although this has no effect classically, 
we expect that this is observable quantum mechanically 
like the Berry phase. See, for example, \cite{berry} for a discussion of 
Berry phase associated with branes in string theory.

Next let us include a (2,0) scalar field $\phi$. As above we assume  
that it satisfies the 
BPS equation \eq{ab-BPS2}, then the BPS equation is satisfied automatically
if we identify $\phi^{(0)} = B^{(0)}_{tw}$. As a result, we have
\be \label{eq:scalarprofile}
\phi^{(0) \;a}
=  -u
\left( \coth(v r) - \frac{1}{v r} \right) \frac{x^a }{r}.
\ee
where
\be
u:= \frac{v}{2 \pi R c}
\ee
set the scale of the vev of $\phi^{(0)}$ at large $r$ since
we can say $\phi^{(0)} \to -\frac{|v|}{2\pi R c} 
x^a T^a/r$ as $r \to \infty$.
In addition, one can define a $U(1)$ projection onto $\phi^{(0)}$.  
This allows us to define the charges 
\be
\begin{split}
P = Q &=\int_{S^1 \times S^2} H^a \phih^a\\
 &=
\mp \int dx^5dS_k\hlf\e_{ijk}\lrbrk{\frac{1}{2\pi Rc}F_{ij}^a \frac{x^a}{r}
+(\text{KK})}\\
 &= - \frac{4\pi}{|c|},
\end{split}
\ee
where 
the $-$ $(+)$ sign in the second equation above corresponds to
the case $c>0$ $(c<0)$; and the term $(\text{KK})$ stands for 
the KK modes and their contribution to the
charges is zero. 
Substituting \eq{constant-c}, we find that the solution is 
self-dual and carries the charges $P=Q=-\sqrt{\pi}$. 
Physically one can identified this self-dual  
string with the uncompactified 
one obtained in the previous section and so they carry the same
charges. 

The scalar profile of (\ref{eq:scalarprofile}) 
is plotted in figure \ref{profile}, 
for two compactification radius $R = 1$ and $R=4$ and a fixed vev
$u=-0.5$. 
One may  compare our results to the scalar profile in \cite{LCCF}. 
In this work, a modified Nahm's equation for the scalar field was conjectured. 
However  unlike the ordinary Nahm's equation where
one can obtain the non-abelian Yang-Mills gauge field at the same time, 
it is not clear how one might obtain 
the corresponding non-abelian tensor gauge field from the modified Nahm's 
equation and the proposal still needed to be completed. 
Nevertheless, qualitatively 
their scalar profile  is  similar to ours.
\FIGURE{
\epsfig{file=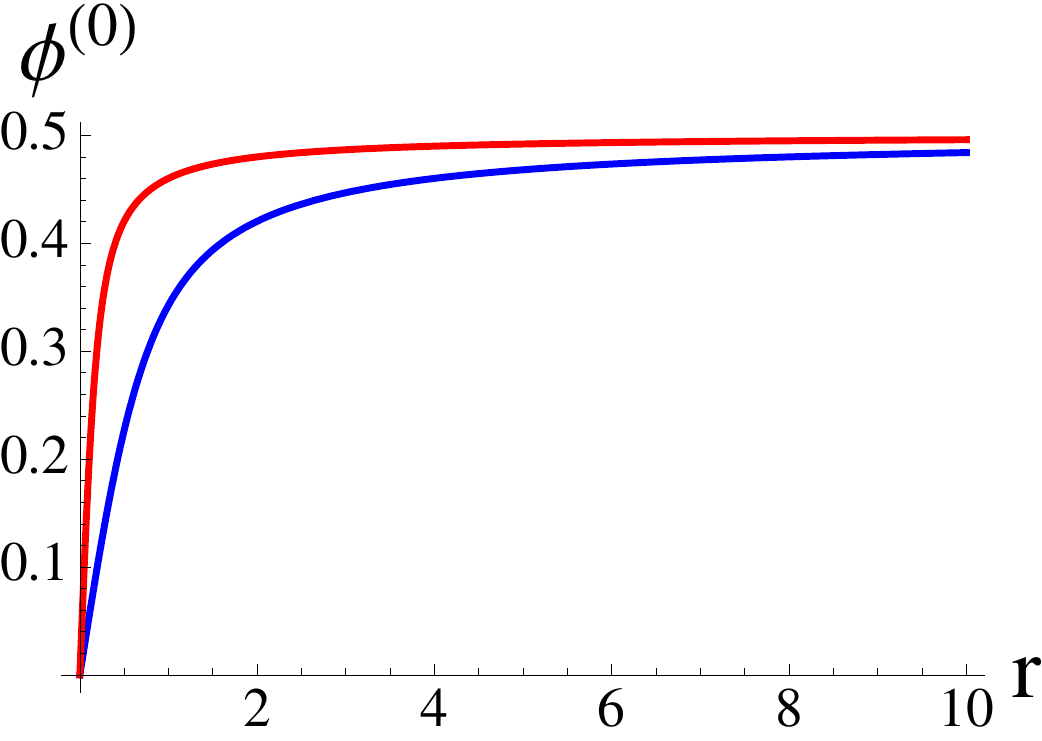,height=6.5cm}
\caption{Scalar Profile. The red curve corresponds to $R=4$
and the blue one to $R = 1$}
\label{profile}
}

\section{Discussions}

In this paper we have constructed the non-abelian string
solutions of the
non-abelian 5-brane theory constructed in \cite{CK}, for both 
uncompactified and compactified spacetime. The string 
solution in non-compact spacetime is supported by a non-abelian Wu-Yang 
monopole, while the string  solution in compact spacetime is 
supported by a non-abelian 't Hooft-Polyakov monopole.
We showed how these solutions can be embedded in the (2,0) supersymmetric theory 
by including a single scalar field obeying a first order 
BPS equation. Although we don't have
the full (2,0) supersymmetric construction yet, we argued that it is 
the correct BPS equation of the (2,0) theory since  it solves 
the equation of motion, and moreover it can be derived from the most natural
form of the supersymmetry transformation law in the non-abelian (2,0) theory.
These string solutions carry self-dual charges and has infinite tension arising 
from the scalar profile which corresponds to having 
a M2-brane spike on the M5-branes system.
These properties are consistent with what one expects 
for the non-abelian
self-dual strings living on a system of two M5-branes. Hence 
the results we obtained 
provide further support that the non-abelian theory 
constructed in \cite{CK} describes the gauge sector of 
a system of multiple M5-branes. 
Needless to say, 
it is of utmost importance to obtain the supersymmetric completion of 
the bosonic theory \cite{CK}. This is under investigation.

We have constructed a non-abelian self-dual string solution with unit
charge. In the M-theory picture, it is possible to have 
non-abelian self-dual strings with higher charges.
It will be interesting to construct them as well. The employment of
multi-monopole seems appropriate, see for
example \cite{shnir,mono1,mono2} for a review. 
It would also be interesting to explore the possible loop space or twistor
interpretation \cite{saemann} of our self-dual string solution.

It is also hoped that the self-dual string solution constructed here 
could provide further insights into the understanding of 
the $N^3$ entropy growth 
of the multiple M5-branes system \cite{m5-S}. 
Recent progress on this problem has been
achieved in \cite{lee,sethi}. 

As advocated in \cite{CS,CG}, just as in the D-branes case where
Lie bracket which define the gauge symmetries for multiple D-branes captures
the noncommutative geometry of a single D-brane 
in the presence of a large NSNS $B$-field, it is possible that 
the gauge symmetry for multiple M5-branes could 
also capture the structure of the quantum geometry of a single M5-branes
in the presence of a large $C$-field.
Given the dynamical evidence we presented 
in this paper, we believe that 
the non-abelian tensor gauge theory of \cite{CK} does describe the gauge sector
of multiple M5-branes. It is thus interesting to try to understand how 
the gauge symmetry of the non-abelian theory \cite{CK} could describe 
the quantum Nambu geometry derived in \cite{CG} for a M5-brane in a 
large $C$-field. An encouraging sign is that both are described in terms 
of ordinary commutator.

\section*{Acknowledgements}

It is a pleasure to thank Anirban Basu, 
David Berman, Neil Lambert,  George Papadopoulos, 
Douglas Smith, Paul Sutcliffe and Martin Wolf for useful conversations
and discussions.
CSC would like to thank 
the Isaac Newton Institute for Mathematical Sciences for hospitality
during the workshop 
``Mathematics and Applications of Branes in String and M-theory''
where part of the work was undertaken.
CSC is supported in part by the STFC Consolidated Grant
ST/J000426/1. 
PV is supported by a Durham Doctoral Studentship 
and by a DPST Scholarship from the Royal Thai Government.

\end{document}